\documentstyle[12pt]{article}
\def\lo{\langle 0 |}
\def\ro{ | 0 \rangle }

 \def\rro{| \rho(q) \rangle }
 \def\fro{ f_{\rho}}
 \def\mro{m_{\rho}}
 \def\emu{\varepsilon_{\mu}^{(\lambda)} }
 \def\enu{\varepsilon_{\nu}^{(\lambda)} }
\def\ealpha{\varepsilon_{\alpha}^{(\lambda)} }
 
\def\gmmu{\gamma _{\mu}}

\def\gmf{\gamma _{5}}

\def\la{\langle }
\def\ra{ \rangle }

\def\sp{ \Sigma_{\ast \cdot} }
\newcommand{\beq}{\begin{equation}}
\newcommand{\eeq}{\end{equation}}
\newcommand{\bea}{\begin{eqnarray}}
\newcommand{\eea}{\end{eqnarray}}

\begin{document}
\renewcommand{\thefootnote}{\fnsymbol{footnote}}
                                        \begin{titlepage}
\begin{flushright}
hep-ph/9704265
\end{flushright}
\vskip1.8cm
\begin{center}
{\LARGE
Conformal symmetry on the light cone  \\
 \vskip0.7cm
and nonleading twist distribution amplitudes \\
\vskip1.0cm
 of massive vector meson }

\vskip1.5cm
 {\Large Igor~Halperin} 
\vskip0.2cm
        Physics and Astronomy Department \\
        University of British Columbia \\
        6224 Agriculture Road, Vancouver, BC V6T 1Z1, Canada \\  
     {\small e-mail: 
higor@physics.ubc.ca }

\vskip1.5cm
{\Large Abstract:\\}
\end{center}
\parbox[t]{\textwidth}{
A complete set of asymptotic three particle light cone 
distribution amplitudes of twist 3 and 4 for a transversely
polarized massive vector meson built out of 
massless current quarks is constructed.
The method used is based on a modified conformal
projectors technique which allows to handle kinematical
corrections due to a finite hadron mass. Consequences 
of our finding for  the $ \rho $-meson hard 
diffractive electroproduction and  $ \gamma \rho 
\pi $ form factor are discussed. Our results may 
imply a breakdown of OPE for some 
exclusive processes beyond the leading twist level.

 }

\vspace{1.0cm}
\begin{center}
{\em submitted to Phys. Lett. B }
\end{center}
                                    \end{titlepage}

\section{Introduction}

Light cone distribution amplitudes (DA's) of 
hadrons are the 
key ingredient of the modern QCD approach to
hard \cite{Brod} and soft \cite{BH} exclusive
, diffractive
\cite{BFGMS} and deep inelastic scattering \cite{BJ}
processes. They provide the universal 
non-perturbative input
in physical amplitudes where the only dependence on 
a particular process enters as a normalization point
of a corresponding DA. Respectively,
they have to be addressed using non-perturbative methods.
The approach of the QCD sum rules pioneered for the 
study of light cone DA's
by Chernyak
and Zhitnitsky more than a decade ago \cite{CZ} 
hitherto 
remains the most popular and reliable among methods   
intended to handle this problem.

While a good deal of information was obtained 
on leading twist DA's for different hadrons, 
relatively a little has been known about nonleading 
twist DA's except for the case of pion. A systematic 
approach to the study of higher twist DA's  
was suggested in \cite{BF} where a complete set of 
light cone DA's of twist 3 and 4 for the pion has
been constructed and studied in detail. The analysis 
of Ref.\cite{BF} was heavily based on group-theoretical
methods exploiting the conformal symmetry which holds
in QCD at the one-loop level \cite{Mak,Ohr,BB}. The 
conformal invariance has proved useful for the study
of leading twist DA's \cite{BL} where it allows to 
diagonalize the mixing matrix at one loop. The use
of the conformal symmetry becomes even more crucial
in the analysis of nonleading twist DA's where methods 
based on direct calculations of suitable correlation 
functions turn out either too complicated or erroneous
 \cite{BF}.

The aim of this letter is to extent the conformal
 group
methods of Ref.\cite{BF} to a study of nonleading
 twist
three particle light cone DA's for a massive vector 
meson built out of massless current quarks
\footnote{See Ref.\cite{BalBr} for a recent 
re-analysis 
of leading twist $\rho$-meson DA's.}.
Fortunately, it requires only a slight
 modification of 
the technique developed in \cite{BF} to handle
 kinematical
corrections due to a non-vanishing hadron mass. 
This method yields quite a different classification 
and new results for twist 4 DA's in comparison to
 the analysis
of Ref.\cite{CZ}. Our
interest in this problem is mainly due to its 
utility for
the diffractive vector meson electroproduction 
at HERA
\cite{HERA} (see \cite{Fr} for updated review of
 the 
experimental and theoretical state of affair there).
 Recently
the perturbative QCD approach of Ref.\cite{BFGMS} 
to the diffractive 
electroproduction of a longitudinally polarized 
vector meson
was reformulated directly in the coordinate space
 in the 
language of Operator Product Expansion 
(OPE) \cite{me}. 
The method
of Ref.\cite{me} easily incorporates an asymmetry 
of the gluon
distribution in the nucleon which is important for
 the whole
problem \cite{Rad}. Moreover, higher twist DA's
 appear quite 
naturally in this approach. Though this paper
 deals with a 
transversely polarized vector meson for which 
the method of 
\cite{me} must be somewhat modified, as a 
by-product of our 
study we show that there are no three-particle 
DA's in the 
diffractive amplitude of Ref.\cite{BFGMS,me} (
see the end of Sect.3).
 Among other 
potential applications one can 
mention the 
light cone QCD sum rule approach \cite{BF1} to 
semileptonic
 and radiative
B-decays \cite{ABS} and (non-)factorization in
 non-leptonic 
B-decays \cite{non}. Besides these practical issues,
our results may have some conceptual importance
as they seem to imply a breakdown of OPE for some
exclusive processes 
at the nonleading twist level.
The letter is organized as follows. In Sect.2 
we review the 
conformal group technique of Ref.\cite{BF}. Its 
modification for 
the case of interest is presented in Sect.3 where we 
construct 
a complete set of asymptotic three particle  DA's for 
a transversely polarized massive vector meson. 
Sect.4 contains
a comparison with previous analyses and discussion of  
possible consequences of our results.  
 
\section{ Conformal symmetry on the light cone 
and asymptotic distribution 
amplitudes}

The aim of  this section is to remind the algebra 
of the so-called collinear 
conformal group obtained by the reduction of the
 full conformal group to
fields varying on the light cone $ x^2 = 0 $ and 
establish a projection operator 
technique and general formulae 
for asymptotic distribution amplitudes with arbitrary
 number of constituents.

 The algebra of the conformal group O(4,2) is obtained 
by adding the generators $ D $ of dilatations 
and $ K_{\mu} $
 of special conformal transformations 
 to the generators $ P_{\mu} $ and $ M_{\mu \nu} $
of the Poincare group. They form the algebra
\bea 
\label{1}
\left[ D, K_{\mu} \right]  &=& i K_{\mu}  \; \;, \; \; 
 \left[ D, P_{\mu}
\right] = -i P_{\mu} 
\nonumber \\
\left[ D, M_{\mu \nu} \right] &=& 0 \; \; , \; \; 
\left[ K_{\mu}, K_{\nu} \right] = 0 \\
\left[ K_{\mu}, P_{\nu} \right] &=& -2i ( g_{\mu \nu} D 
+ M_{\mu \nu}) \nonumber \\
\left[ K_{\alpha}, M_{\mu \nu} \right] &=& i 
( g_{\alpha \mu} K_{\nu} - g_{\alpha \nu}
K_{\mu})    \nonumber
\eea
in addition to the commutation relations 
of the Poincare algebra.
These generators act on a field $ \Phi(x) $ with 
the spin $ s $ and canonical 
dimension $ l $ as follows
\bea
\label{2}
\left[P_{\mu}, \Phi(x) \right] &=& - i \partial_{\mu}
 \Phi(x) \nonumber \\
\left[ M_{\mu \nu}, \Phi(x) \right] &=& [i (x_{\nu} 
\partial_{\mu} - x_{\mu} \partial_{
 \nu}) - \Sigma_{\mu \nu} ] \Phi(x)  \nonumber \\
\left[ D, \Phi(x) \right] &=& -i ( x_{\alpha} 
\partial_{\alpha} + l ) \Phi(x)  \\
\left[ K_{\mu}, \Phi(x) \right] &=& -i ( 2 x_{\mu}
 x_{\alpha} 
\partial_{\alpha}
- x^2 \partial_{\mu} + 2 x_{\mu} l - 2 i x_{\nu}
 \Sigma_{\mu \nu} ) 
\Phi(x)  \nonumber
\eea 
where $ \Sigma_{\mu \nu} $ is the spin generator 
\beq
\label{3} 
\Sigma_{\mu \nu} \Psi = \frac{1}{2} \sigma_{\mu
 \nu} \Psi \;, \; 
\Sigma_{\mu \nu} G_{\alpha \beta} = i ( 
\delta_{\mu \alpha} 
G_{\nu \beta} - \delta_{\nu \alpha} G_{\mu 
\beta} ) 
- (\alpha \leftrightarrow \beta)
\eeq
 For an analysis of the light cone hadron
 distribution amplitudes we 
only need to consider the action of the conformal
 generators (\ref{1})
on a field $ \Phi(ux) $ varying on the light cone
 $ x^2 =0 $. It is convenient to simplify
the algebra by introducing two 
light like vectors
$ x_{\mu } , \bar{x}_{\mu} $ with $ x^2 = 
\bar{x}^2 = 0 \; , \; (x \bar{x}) 
= 1 $ and defining projections on these vectors
 $ \gamma_{\cdot} = 
\gmmu x_{\mu} \; , \; \gamma_{\ast} = \gmmu 
\bar{x}_{\mu} $ etc. 
Then the only nontrivial transformations
 on the field $ 
\Phi(ux) $ are given by the 
components $ P_{\cdot} \;, \; D \; ,\; 
 M_{\ast \cdot} \; , \; K_{\ast} $. 
These generators form the algebra of the so-called 
collinear conformal 
subgroup $ SO(2,1) \simeq SU(1,1) \simeq 
SL_{2}(R) $ of projective 
transformations on the line. Its  
algebra reads
\footnote{Note the ``wrong" sign of the commutator 
$ [ J_{+} , J_{-} ] $ in comparison to the  
rotation group SO(3).} 
\bea
\label{4}
\left[ J_{+} , J_{-} \right] &=& - 2 J_{3} 
\; \; , \; \, 
\left[ J_{3}, J_{\pm} \right] = \pm J_{\pm} 
 \nonumber \\
\left[ J^2 , J_{i} \right] &=& 0 \; \; , \; \;
 \left[
E, J_{i} \right] = 0 
\eea
where we defined the operators $ J_{i} $ as linear 
combinations of 
the conformal operators
\bea
\label{5}
J_{+} &=& J_{1} + i J_{2} = \frac{i}{\sqrt{2}}
 P_{\cdot} \; , \; 
J_{-} = J_{1} - i J_{2} =  \frac{i}{\sqrt{2}}
 K_{\ast} \nonumber \\
J_{3} &=& \frac{i}{2} ( D + M_{\ast \cdot}) \; 
,\; E = \frac{i}{2}
( D- M_{\ast \cdot} )  \\
J^2 &=& J_{3}^2 - J_{1}^2 - J_{2}^2 = J_{3}^2
 - J_{3} - J_{+} J_{-}
\nonumber
\eea
Here two observations are in order. First, the
 light like vectors 
$ x_{\mu}\;,
\; \bar{x}_{\mu} $ introduced above are quite
 arbitrary at this stage. 
However, they have to be built up from physical 
vectors characteristic
to a meson in order to get informative constraints
 for distribution 
amplitudes. Ways to do this will be described in
 the next section. Second,
we note that the use of the collinear conformal group
 requires fixing a projection of 
the Lorentz spin of the field $ \Phi(ux) $ onto the 
line $ x_{\mu} $
\beq
\label{6}
\Sigma_{\ast \cdot} \Phi(ux) = i s \Phi(x) 
\eeq
In this case the field $ \Phi(ux) $ is an eigenstate
 of the Casimir 
operator $ J^2 $ :
\beq
\label{7}
\left[ J^2 , \Phi(ux) \right] = j ( j -1 ) \Phi(ux) 
\eeq
where $ j = 1/2 (l + s) $ is the conformal spin of 
the field $ \Phi(ux)$.
The fixed spin components of the quark $ \Psi(ux) $ 
and gluon 
$ G_{\mu \nu}
(ux) $ fields can be readily constructed as follows. 
The quark field contains 
two components with the spin projections $ s = 
\pm 1/2 $ which can 
be selected by the projection operators $ \gamma_{\cdot}$
 and $ \gamma_{\ast}
$ :
\bea
\label{8}
\Sigma_{\ast \cdot} \gamma_{\cdot} \Psi &=& \frac{1}{2} 
\sigma_{\ast \cdot} \gamma_{\cdot} \Psi = + i\frac{1}{2}
\gamma_{\cdot} \Psi       \nonumber \\
\Sigma_{\ast \cdot} \gamma_{\ast} \Psi &=& \frac{1}{2} 
\sigma_{\ast \cdot} \gamma_{\ast} \Psi = - i\frac{1}{2}
\gamma_{\ast} \Psi 
\eea
For the gluon field the relevant components with 
the spin 
projections 
$ s = 0, \pm 1 $ are the following
\beq
\label{9}
\sp G_{\cdot \perp} = +i G_{\cdot \perp} \; , \; 
\sp G_{\ast \perp} = - i G_{\ast \perp} \; , \;
 \sp 
G_{\ast \cdot}
= \sp G_{\perp \perp} = 0 
\eeq
In the last equation the label $ \perp $ stands
 for 
projections on the 
third vector $ e_{\alpha}^{(\perp)} $ which is
 required 
to satisfy 
$ ( \bar{x} e^{(\perp)} ) = 0 $. 

In an analysis of hadron DA's we deal with matrix
 elements 
of the form
\beq
\label{10}
\lo \Phi_{1}(u_{1}x ) \ldots \Phi_{k}(u_{k}x)|
 h(q) \ra =
\int D \alpha e^{-iqx(u_{1} \alpha_{1} + \ldots
 + u_{k} 
\alpha_{k}
) } \phi(\alpha_{1}, \ldots, \alpha_{k} )
\eeq
where $ D \alpha = d \alpha_{1} \cdots \alpha_{k} 
\delta(\sum_{i}
\alpha_{i} - 1 ) $. Here $ h $ is a hadron of
 interest and $
 \phi(\alpha_{1}, \ldots, \alpha_{k} ) $ stands for a 
corresponding 
multiparticle DA. The fields $ \Phi_{i}(u_{i}x) $ 
are arranged 
along the
light cone $ x^2 = 0 $. The light cone conformal
 symmetry 
suggests the 
following way for the analysis of the DA  $
 \phi(\alpha_{1}, \ldots, \alpha_{k} ) $. First,
 we choose 
fixed spin 
components of the fields $  \Phi_{i}(u_{i}x) $ 
according to 
Eqs.(\ref{8}),(\ref{9}). Next, we note that an expansion 
of the 
tensor product $  \Phi_{1}(u_{1}x ) \ldots 
\Phi_{k}(u_{k}x) $ 
into irreducible representations of the collinear 
conformal group
induces a respective decomposition for the function $  
\phi(\alpha_{1}, \ldots, \alpha_{k} ) $ defined on 
the simplex 
$ \alpha_{1} + \ldots + \alpha_{k} = 1 $. Therefore 
the evaluation
of the DA $ \phi(\alpha_{1}, \ldots, \alpha_{k} ) $ 
amounts 
to a calculation of corresponding Clebsch-Gordan 
coefficients 
in an expansions of a tensor product into irreducible 
representations 
with a fixed value of the total conformal spin. An 
asymptotic 
DA is then defined as a representation with 
the lowest
conformal spin which is $ j_{min} = j_{1} + 
\ldots j_{k} $. 
A general formula for the asymptotic form 
of arbitrary multiparticle DA has been found in
 Ref.\cite{BF} 
exploiting the fact that the collinear conformal 
group (\ref{4}) is 
identical to the Lorentz group in (2+1) dimensions.
 It reads
\beq
\label{11}
\phi_{as}(\alpha_{1}, \ldots, \alpha_{k}) = 
\frac{\Gamma(2 j_{1} + \ldots + 2 j_{k} )}{ 
\Gamma(2j_{1}) \ldots 
\Gamma(2 j_{k})} \alpha_{1}^{2 j_{1} -1} \ldots 
\alpha_{k}^{2 j_{k}
-1}  \; \; , \; \; \int D \alpha \phi_{as} = 1 
\eeq
Irreducible representations with higher spins
 are obtained 
by the multiplications of (\ref{11}) by 
mutually orthogonal polynomials 
of  
dimensions $ j - \sum_{i} j_{i} $ and correspond
 to conformal 
corrections to the asymptotic multiparticle DA 
 $ \phi(\alpha_{1}, \ldots, \alpha_{k} ) $ 
which serves as the weight function 
at the space of conformal polynomials. These
 conformal corrections will not be 
discussed in the present paper.     
 
Going over to interacting fields we insert the
 gauge factors 
$ [ux,vx] =  \\ 
P \exp[ig \int_{v}^{u}
 d t x_{\mu} A_{\mu}(tx) ] $ in between the
 field
operators in Eq.(\ref{10}) which is equivalent
 to the 
substitution of ordinary derivatives by covariant
 ones 
in the language of local conformal 
operators. Then contributions
of different conformal spins in DA do not mix under 
renormalization to the one-loop accuracy \cite{Mak}.
This property is sufficient to construct an expansion
of the leading twist DA over a series of polynomials 
with multiplicatively renormalizable coefficients, as 
in this case all representations are nondegenerate.
Moreover, anomalous dimensions are ordered with the 
conformal spin for the leading twist.  
For higher twist DA's irreducible representations with 
sufficiently high spins are degenerate and conformal 
operators do not completely diagonalize the mixing
matrix which only becomes ``conformal block-diagonal".
Still, anomalous dimensions are 
typically lowest for operators of minimal conformal 
spin. These operators are therefore leading ones 
as $ Q^2 \rightarrow \infty $ and thus in this 
case the definition
(\ref{11}) is justified.
  
\section{Distribution amplitudes of a massive 
vector meson}

 In this section we explain how the general analysis 
of the 
proceeding section works in the study of 
three-particle
quark-antiquark-gluon DA's of twist 3 and 4 of a 
massive vector 
meson. We consider a task of taking into account
 kinematical 
corrections due to a nonvanishing meson mass while 
the 
constituents of a meson are assumed to be massless. 
Thus, our 
results apply directly to the $ \rho - , 
\omega- , 
\phi-$ mesons (SU(3) limit is implied)
and also can be used for the 
charmonium in
 situations
where one can neglect the c-quark mass. For 
definiteness, we 
consider the case of the charged $
 \rho-$meson, while
modifications to be done for other mesons are trivial.  
 
 We start with a set of three particle DA's of
 twist 3 and 4 defined by the matrix elements 
of the following non-local 
light cone string operators
\footnote{We assume the Fock-Schwinger gauge $ x_{\mu}
A_{\mu}(x) = 0 $. In a general case path-ordered 
gauge factors between the fields in (\ref{12},\ref{13})
are implied, as required by the gauge invariance.}
\bea
\label{12}
\lo \bar{d}(-x) \gamma_{\alpha} i g G_{\mu
 \nu} (vx) u(x) \rro =
 q_{\alpha}( \emu q_{\nu} - \enu q_{\mu} )
 \int D (\alpha_{i} qx) 
 \Phi^{(1)}(\alpha_{i} ) \nonumber \\
+ \ealpha(q_{\mu} x_{\nu}- q_{\nu} x_{\mu})
\frac{m^2}{qx}  \int D (\alpha_{i} qx) 
 \Phi^{(2)}(\alpha_{i} )  \\
+ ( \emu  \delta_{\alpha \nu} - 
  \enu  \delta_{\alpha \mu} )  
 m^2 \int D (\alpha_{i}qx)  
 \Phi^{(3)}(\alpha_{i} )     \nonumber
\eea
\bea
\label{13}
\lo \bar{d}(-x) \gamma_{\alpha} \gmf g 
\tilde{G}_{\mu \nu} (vx) u(x) \rro =
 q_{\alpha}( \emu q_{\nu} - \enu q_{\mu} ) 
 \int D (\alpha_{i}qx) 
 \Psi^{(1)}(\alpha_{i} ) \nonumber \\
+ \ealpha(q_{\mu} x_{\nu}- q_{\nu} x_{\mu})
\frac{m^2}{qx}  \int D (\alpha_{i} qx) 
 \Psi^{(2)}(\alpha_{i} )  \\
+ ( \emu  \delta_{\alpha \nu} - 
  \enu  \delta_{\alpha \mu} )  
 m^2 \int D (\alpha_{i}qx)   
\Psi^{(3)}(\alpha_{i} )     \nonumber
\eea
 Here 
\beq
\int D(\alpha_{i}qx) = \int d \alpha_{1} 
d\alpha_{2} d \alpha_{3} \delta(
1-\alpha_{1} -\alpha_{2} -\alpha_{3})
 e^{-iqx(\alpha_{1} -\alpha_{2} + v 
\alpha_{3} )}
 \eeq
and $ m^2 $ stands for the meson mass.
One can check that the DA's  $  
\Phi^{(1)} , \Psi^{(1)} $ corespond to  
operators of twist 3, while the 
DA's $ \Phi^{(2)}, \Phi^{(3)}, 
 \Psi^{(2)}, \Psi^{(3)}
$ contain contributions of 
operators of both twist 3 and 4.   
This can be seen using the 
operator identity \cite{BB}
\bea
\label{tw3}
\left[ \bar{\psi}(-x) \gamma_{\alpha} \gmf 
g \tilde{G}_{\mu \cdot}(vx) \psi(x) \right]^{tw- 3} 
= \int_{0}^{1}du \frac{1+u^{2}}{2} \frac{\partial}{
\partial x_{\alpha}} \bar{\psi}(-ux) \gamma_{\cdot}
\gmf g \tilde{G}_{\mu \cdot}(uvx) \psi(ux) \nonumber \\
+ \int_{0}^{1}du \frac{1- u^{2}}{2} \frac{\partial}{
\partial x_{\mu}} \bar{\psi}(-ux) \gamma_{\cdot}
\gmf g \tilde{G}_{\alpha \cdot}(uvx) \psi(ux)
\eea
Multiplying (\ref{tw3}) by $ x_{\alpha} $, integrating 
by parts and taking matrix elements in accord with 
(\ref{13}), one arrives at identities for the 
DA's   $  
\Phi^{(1)} , \Psi^{(1)} $.
We further note that  the $\rho$-meson
in Eqs.(\ref{12}),(\ref{13}) can only 
 be transversely polarized. Indeed, the total
helicity of the quark-antiquark pair is zero
for the vector or axial channel. Therefore, in order
to have a longitudinally polarized meson
the gluon must also be longitudinal. This is,
however, forbidden by the gauge invariance as
the hadron constituents in DA are nearly
 on-shell. This 
argument has been known for some time 
\cite{Chernyak}.
Still, as it has a somewhat hand waving flavor (the
meaning of being ``nearly on-shell" is not quite
clear), we have formally checked this statement 
by allowing for longitudinal structures in (\ref{12},
\ref{13}) and respective extending a set of conformal 
constraints (see Eqs.(\ref{19}- \ref{21}) below). 
A corresponding system of equations turns out 
inconsistent, which means that the longitudinal 
polarization in (\ref{12},\ref{13}) is not possible.
It is then easy to see that the Lorentz structure 
displayed in Eqs.(\ref{12}),(\ref{13}) is the most
 general
decomposition to the O(twist-5) accuracy.
In the early analysis of Ref.\cite{CZ} only 
the twist 3 structures $  
\Phi^{(1)} ,
 \Psi^{(1)} $ were retained on the 
ground of the fact that they correspond to leading 
power corrections in exclusive amplitudes. They have
further been analysed within the QCD sum rules technique. 
We are going to extent the analysis of the DA's 
(\ref{12},\ref{13}) by a completely different method which
is also able to fix the ``twist 4"
\footnote{Here we put the quotation marks in 
order to remind that these DA's give rise 
to further power corrections in exclusive 
amplitudes in comparison to those produced 
by the DA's  $  
\Phi^{(1)} ,
 \Psi^{(1)} $. One has, however, to bear in mind
that in the operatorial sense they contain 
contributions of twist 3, as can be seen from Eqs.
(\ref{17}-\ref{19}).} DA's $ 
 \Phi^{(2)}, \Phi^{(3)},  \Psi^{(2)}, \Psi^{(3)}
$. Besides of obvious practical interest, we find that 
our new results for the ``twist 4" DA's may have some 
principal importance as well. The point is that,
as will be shown below, there is a very nontrivial
interplay between the expansions in twist and conformal
spin in the problem at hand. Non-leading ``twist 4" DA's 
have lower
in comparison to twist 3 DA's conformal spins, i.e. are 
given by polynomials of lower dimensions. This observation
may imply a breakdown of OPE for some processes at the 
nonleading twist level (see Sect.5). In what follows 
we restrict our study to the asymptotic DA's, i.e. work
to the leading order in the conformal spin expansion.
A calculation of first pre-asymptotic corrections is 
both possible and desirable. Unfortunately, it is 
notoriously more difficult in comparison to the pion
case \cite{BF} due to a larger number of possible
Lorentz structures in matrix elements corresponding 
to conformal corrections. 

After these lengthy remarks we proceed to the 
conformal analysis of the DA's (\ref{12}),(\ref{13}).
As has been explained in the previous Sect.2, to 
this end we need to construct two light like vectors
$ x_{\mu} , \bar{x}_{\mu} $. As for the first 
one, it is most conveniently chosen to be just 
the light cone coordinate $ x_{\mu} $ which enters 
the definitions (\ref{12}),(\ref{13}) (this
justifies the identical notations for both). 
A form of the 
second light like vector $ \bar{x}_{\mu} $ is a bit 
less obvious. The choice $ \bar{x}_{\mu} = 
q_{\mu}/(qx) $ which was used in Ref.\cite{BF} for 
the analysis of the pion DA's clearly does not work 
in the case at hand as now $ q^2 \neq 0 $. The correct
form of the vector $ \bar{x}_{\mu} $ for
 a massive vector
meson with arbitrary polarization is given by
\beq
\label{16}
\bar{x}_{\mu} = \frac{q_{\mu} + m \emu}{ qx +
 m (\varepsilon^{(\lambda)} x)} \; \; , 
\; \; \bar{x}^2 = 0
\eeq
supplemented with $ ( \varepsilon^{(\lambda)} 
e^{(\perp)} ) = 0 $. It is easy to check that with
such choice the projection formulae (\ref{8}),(\ref{9})
and the whole analysis of Sect.2 remain valid 
in the massive case. 

 Selecting 
different spin components in Eq.(\ref{12}) according to
(\ref{8}),(\ref{9}),(\ref{16}) we obtain the following 
system
\beq
\label{17}
\lo \bar{d}(-x) \gamma_{\cdot}igG_{\ast \cdot}(vx)
u(x) \rro = - m (qx) \int D(\alpha_{i} qx)  
 \Phi^{(1)} 
\eeq
\beq
\label{18}
\lo \bar{d}(-x) \gamma_{\ast}igG_{\ast \cdot}(vx)
u(x) \rro = \frac{m^3}{ (qx)} \int D(\alpha_{i} qx) 
\left( 
\Phi^{(2)} - \Phi^{(1)}- \Phi^{(3)} \right) 
\eeq
\beq
\label{19}
\lo \bar{d}(-x) \gamma_{\perp}igG_{\perp \ast}(vx)
u(x) \rro = (e^{(\perp)} e^{(\perp)} )
\frac{m^3}{ (qx)} \int D(\alpha_{i} qx) 
\Phi^{(3)}  
\eeq
\beq
\label{20}
\lo \bar{d}(-x) \gamma_{\perp}igG_{\perp \cdot}(vx)
u(x) \rro =   0   \nonumber 
\eeq
plus analogous equations for the set (\ref{13}) which are 
obtained from (\ref{17} - \ref{20}) by the substitution
$ ig G \rightarrow  \gmf g \tilde{G} \; , \; \Phi^{(i)}
\rightarrow \Psi^{(i)} $.
 We have used the fact that for the transversely 
polarized $ \rho$-meson $ (\varepsilon^{(\lambda)}
x) = 0 $ for $ x^2 = 0 $. We will now study separately
each of the Eqs.(\ref{17} -\ref{19}). Let us start 
with the first couple of equations. Using the 
general formula (\ref{11}) we write 
\footnote{Here and in what follows ellipses 
stand for higher 
conformal spin contributions.}
\bea
\label{21}
\lo \bar{d}(-x) \gamma_{\cdot}igG_{\ast \cdot}(vx)
u(x) \rro &=& 120 a_{1} \int D(\alpha_{i} qx)
\alpha_{1} \alpha_{2} \alpha_{3} ( 1 + \ldots)
\nonumber \\
\lo \bar{d}(-x) \gamma_{\cdot} \gmf g 
\tilde{G}_{\ast \cdot}(vx)
u(x) \rro &=& 120 b_{1} \int D(\alpha_{i} qx)
\alpha_{1} \alpha_{2} \alpha_{3} ( 1 + \ldots)
 \eea
We introduce here matrix elements of local operators
\bea
\label{22}
\lo \bar{d} \gamma_{\alpha} i g G_{\mu \nu}
u \rro &=& q_{\alpha} (\emu q_{\nu} - \enu q_{\mu})
A_{1} + ( \emu \delta_{\alpha \nu} - 
\enu \delta_{\alpha \mu}) m^2 A_{2} \nonumber \\
\lo \bar{d} \gamma_{\alpha} \gmf g 
\tilde{G}_{\mu \nu}
u \rro &=& q_{\alpha} (\emu q_{\nu} - \enu q_{\mu})
B_{1} + ( \emu \delta_{\alpha \nu} - 
\enu \delta_{\alpha \mu} ) m^2 B_{2}
\eea
It is clear that $ A_{1} = A_{2} = 0 $ by the G-parity,
while $ B_{1} $ and $ B_{2} $ are both non-zero and 
will be calculated below. It is then easy to 
express the normalization constants $ a_{1}\; , \; b_{1} $ 
of Eq.(\ref{21}) in terms of $ A_{i} $ and $ B_{i} $ of 
(\ref{22}). In this way we find 
\bea
\label{23}
 \Phi^{(1)} 
  &=& 0 \nonumber \\
 \Psi^{(1)} &=& 
120 B_{1} \alpha_{1} \alpha_{2} \alpha_{3} 
\eea
A second couple of constraints is obtained analogously 
from Eq.(\ref{18}) and its counterpart for $ \Psi_{i} $
. It reads
\bea
\label{24}
 \Phi^{(1)} - 
 \Phi^{(2)} + \Phi^{(3)} &=& 0 \nonumber \\
 \Psi^{(1)} - 
 \Psi^{(2)} + \Psi^{(3)}&=& 6 (B_{1} + B_{2}) \alpha_{3} 
\eea
The last Eq.(\ref{19}) needs more care. Note that the 
quark spins in (\ref{19}) are not fixed by the transverse 
gamma matrix. Using the identity
\beq
\label{25}
\gamma_{\perp} = -\frac{1}{2} ( \gamma_{\cdot} 
\gamma_{\perp} \gamma_{\ast} + \gamma_{\ast} 
\gamma_{\perp} \gamma_{\cdot} ) 
\eeq
we  
consider two auxiliary DA's with fixed 
projections 
of the spins of quarks on 
the line $ x_{\mu} $ :
\bea
\label{26}
\lo \bar{d}(-x) \gamma_{\cdot} \gamma_{\perp} 
\gamma_{\ast} ig G_{\perp \ast}
(vx) u(x) \rro &=& 6 a  \int D(\alpha_{i} qx) 
\alpha_{2}  ( 1 + \cdots)  \nonumber \\
\lo \bar{d}(-x) \gamma_{\ast} \gamma_{\perp}
 \gamma_{\cdot} ig G_{\perp \ast}
(vx) u(x) \rro &=& -6 a \int D(\alpha_{i} qx)
 \alpha_{1}  ( 1 + \cdots)   \; , 
\eea
The normalization factor $ a $ is the same in
 both expressions by the G-parity. One can see
that only the antisymmetric combination of
gamma matrices contributes in Eq.(\ref{26}).
A little algebra then yields $ a = B_{2} m^3 /
 (qx) $. After a similar calculation for the 
axial channel we obtain
\bea
\label{27}
\Phi^{(3)} &=& 3 B_{2} (\alpha_{1} - \alpha_{2})
\nonumber \\
\Psi^{(3)} &=& 3 B_{2} (1-\alpha_{3}) 
\eea 
Using Eqs.(\ref{23}, \ref{24}, \ref{27}) 
we finally arrive at the following 
 expressions for the 
asymptotic DA's :
\bea
\label{28}
\Phi^{(1)} &=& 0  \nonumber \\
\Phi^{(2)} &=& \Phi^{(3)}
= 3 B_{2} (\alpha_{1} - 
\alpha_{2} ) \nonumber \\
\Psi^{(1)} &=& 
 120 B_{1} \alpha_{1} \alpha_{2} 
\alpha_{3}   \\
\Psi^{(2)} &=& 3  B_{2}
(1- 3 \alpha_{3}) + 6 B_{1} \alpha_{3} 
( 20 \alpha_{1} 
\alpha_{2}  -1 )  \nonumber  \\
\Psi^{(3)} &=& 3 B_{2} (1- \alpha_{3}) \nonumber
\eea
Note in passing that a peculiarity of restrictions
imposed by the conformal invariance can be seen 
e.g. in the expression for $ \Psi^{(2)} $
which does not correspond to any simple minded 
homogenious polynomial in $ \alpha_{i} $.
Constants $ B_{1}, B_{2} $ can be calculated 
by the standard QCD sum rules method. In fact, 
their linear combination $ B_{1} + 3 B_{2}
 $ (corresponding to 
the convolution  with $ \delta_{\alpha \nu} $
in Eq.(\ref{22} )) was found a long time ago
\cite{BKol}. In analogy with the method used in 
\cite{BKol} we find  
 convenient to calculate the non-diagonal
 correlation function
\beq
\label{29}
T(q) = i \int dx e^{iqx} \lo T \{ \bar{d}(x) 
 \gamma_{\alpha} \gmf g \tilde{G}_{\mu \nu} u (x) 
\bar{u}(0) \gamma_{\lambda} d(0)  \} \ro
\eeq
The matrix element of interest appears as the lowest
intermediate hadron state contribution to the imaginary 
part
of (\ref{29}). Matching this expression to a QCD 
calculation at $ q^2 \rightarrow - \infty $ we 
obtain two sum rules for parameters $ B_{1} $ and
$ B_{2}$. 
The Borel transformed sum rules look as follows 
(a subtraction of continuum from the perturbative
 term is implied):
\bea
\label{30}
B_{1} &=& \frac{1}{ \mro \fro} e^{\frac{m_{\rho}^2}{M^2}} 
\left[- \frac{\alpha_{s}}{
72 \pi^3} M^4 + \frac{1}{24} \la \frac{\alpha_{s}}{\pi}
 G^2 \ra + \frac{40}{27}
\pi \alpha_{s} \frac{ \la \bar{\psi} \psi \ra^2 }{M^2}
 + \cdots \right] \nonumber \\
B_{2} &=& \frac{1}{ \mro \fro} e^{\frac{m_{\rho}^2}{M^2}}
 \left[- \frac{\alpha_{s}}{
72 \pi^3} M^4 + \frac{1}{24} \la \frac{\alpha_{s}}{\pi}
 G^2 \ra - \frac{24}{27}
\pi \alpha_{s} \frac{ \la \bar{\psi} \psi \ra^2 }{M^2}
 + \cdots \right] 
\eea
Unfortunately, the first of the sum rules in 
(\ref{30}) is 
not stable as the condensate terms enter there 
with the same signs. Therefore we use only the second 
of Eqs.(\ref{30}) to find the number $ B_{2} $ and 
fix the other parameter $ B_{1} $ from the known value 
$ B_{1} + 3 B_{2} \simeq 1.85 \cdot 10^{-2} \; GeV^2 $
which was obtained in Ref.\cite{BKol} from a similar
(stable) sum rule written directly for this 
combination.   
Numerically this procedure  yields
\beq
\label{31}
B_{1} \simeq 0.8 \cdot 10^{-2} \; GeV^2 \; , \;  
B_{2} \simeq 0.35 \cdot 10^{-2} \; GeV^2   
\eeq
These numbers provide normalizations to the
complete system of asymptotic
DA's (\ref{28}). 
A comparison between our results (\ref{28})
and the set of DA's suggested in Ref.\cite{CZ}
will be carried out in the next section.
Here we would like to end up the discussion 
of the $ \rho$-meson DA's with a remark that
 there exists one more series of DA's
generated by the matrix element
\beq
\label{dop}
\lo \bar{d}(-x) \sigma_{\mu \nu} ig G_{\alpha \beta}
(vx) u(x) \rro
\eeq
which corresponds to a longitudinally polarized vector
meson \cite{Chernyak} as there the total quark helicity
is $ \pm 1 $ and the gluon helicity is $ \mp 1 $. 
These DA's cannot contribute the diffractive amplitude
of \cite{BFGMS,me} owing to odd number of gamma
matrices in a trace. This justifies the claim made in 
Introduction on the absence of three particle DA's in 
the diffractive electroproduction amplitude of 
Ref.\cite{BFGMS,me}

\section{Discussions}

Our results (\ref{28}) should be confronted with
the alternative 
set of DA's suggested in Ref.\cite{CZ} :
\bea
\label{32}
\Phi^{(1)} &=& 7! f_{3 \rho}^{V}
(\alpha_{1} - \alpha_{2}) \alpha_{1} \alpha_{2}
\alpha_{3}^2 \;, \; \int D \alpha (\alpha_{1} - 
\alpha_{2}) \Phi_{\parallel}^{(1)} =  
 f_{3 \rho}^{V} \simeq 0.25 \cdot
10^{-2} \; GeV^2  \nonumber \\
\Psi^{(1)} &=& 360 f_{3 \rho}^{A}
 \alpha_{1} \alpha_{2}
\alpha_{3}^2 \;, \; \int D \alpha 
 \Psi_{\parallel}^{(1)} =  
 f_{3 \rho}^{A} \simeq 0.6 \cdot
10^{-2} \; GeV^2  
\eea
As the DA's $ \Phi^{(1)} $ in (\ref{28}) and (\ref{32}) 
are incomparable (any nontrivial  $ \Phi^{(1)} $ 
is pre-asymptotic within a systematic expansion
in the conformal spin and thus is beyond our 
asymptotic accuracy), we concentrate on a comparison 
between the DA's  $ \Psi^{(1)} $. First we note that 
the normalization parameter $ B_{1} $ of the DA's 
 $ \Psi^{(1)} $ in (\ref{28}) is consistent within a 
typical for the QCD sum rules method accuracy with 
the normalization $  f_{3 \rho}^{A} $ in (\ref{32}).
However, the functional forms of $ \Psi^{(1)} $
in (\ref{28}) and (\ref{32}) are different by an 
extra power of $ \alpha_{3} $. A source of this 
discrepancy is not quite clear to us. It can be 
due to different definitions of twist in 
this paper and Ref.\cite{CZ}. Our  
$ \Psi^{(1)} $
corresponds to matrix elements of twist 3  
operators, while 
the authors of \cite{CZ} have in fact singled out 
symmetric parts of operators without subtraction
of traces, as just these quantities determine 
power corrections in exclusive amplitudes.
Another possible and related reason is that 
when (as in Ref.\cite{CZ})
the DA is calculated by studying a correlation
function of currents with derivatives together 
with a symmetrization procedure, the imaginary part
becomes a combination of contributions of different 
resonances in different Lorentz structures. In this case 
the DA calculated from the asymptotic loop may correspond 
to some ``average" hadron (i.e. an admixture of a
hypothetic exotic meson with $ J^{PC} = 0^{+ -} $ 
\cite{BKol} is
 not ruled out).
We also note that from the viewpoint of the conformal
spin expansion the polynomial $ \alpha_{1} \alpha_{2}
\alpha_{3}^{2} $ could arise from the next-to-leading
term $ \alpha_{1} \alpha_{2}
\alpha_{3} ( 1- 3 
\alpha_{3} ) $ with $ j = 7/2 $, while our 
asymptotic $ \Psi^{(1)}
$ has $ j = 3 $ \footnote{A first possible non-zero 
term in the DA $ \Phi^{(1)} $ corresponds to the 
polynomial $  ( \alpha_{1} - \alpha_{2} )
\alpha_{1} \alpha_{2}
\alpha_{3} $ with $ j = 7/2 $. The same power of
 $ \alpha_{3}
$ seems to be missing there as well.}.
A coefficient in front of this conformal correction
is fixed by the $ \rho $- meson matrix element of the
 $ j = 7/2 $ 
operator $ \bar{d} \gamma_{\cdot} \gmf 
g D_{\cdot} \tilde{G}_{\ast \cdot} u $. On the other 
hand, it can be readily seen that this matrix element
is exactly zero for the transverse polarization
of the $ \rho$-meson. Therefore, it seems unlikely to 
have a multiplicatively renormalizable combination
of conformal operators giving rise to the combination
 $ \alpha_{1} \alpha_{2}
\alpha_{3}^{2} $ and having an anomalous dimension
lower than that of the $ j= 3 $ operator   $ \bar{d} 
\gamma_{\cdot} \gmf 
g \tilde{G}_{\ast \cdot} u $ which corresponds to 
the polynomial  $ \alpha_{1} \alpha_{2}
\alpha_{3} $.
To summarize, we see a few possible explanations for the 
difference by a power of $ \alpha_{3} $ between the DA's
$ \Psi^{(1)} $ in (\ref{28}) and (\ref{32}), but for 
its precise identification a further analysis is needed.
While the choice (\ref{32})
may be ultimately true for the pQCD
method of Ref.\cite{CZ}, it seems 
that it is the DA  $ \Psi^{(1)} $ of Eq.(\ref{28})
that has to be used in the light cone QCD approach to 
``soft" exclusive processes \cite{BH}, since there one 
deals directly with the matrix elements (\ref{12},\ref{13}).

Now we would like to discuss the non-leading 
``twist 4" DA's
$ \Phi^{(2)}, \Phi^{(3)},  \Psi^{(2)}, \Psi^{(3)}
$ in (\ref{28}) which hitherto have not been 
considered in the literature. Somewhat surprisingly,
we have found that though these DA's 
correspond to further power corrections
in comparison to those produced by the twist 3 DA's
$ \Phi^{(1)}, \Psi^{(1)} $, their conformal spins 
are lower ($ j = 2 $ for  $ \Phi^{(2)}, \Phi^{(3)}, 
$ and $ \Psi^{(3)} $, while $  \Psi^{(2)} $ is 
a combination of $ j = 2 $ and $ j = 3 $ pieces). 
Respectively, they are given by polynomials of lower 
dimensions than the asymptotic DA $ \Psi^{(1)} $ 
in (\ref{28}). On the other hand, we remind that 
typically DA's enter exclusive amplitudes through 
some integrals like $
\int \Phi(\alpha_{i})/ \alpha_{3}^2 $, cf. Ref.\cite{CZ}.
Powers of $ \alpha_{i} $ in the denominator originate 
from quark and gluon propagators and are therefore
the same in terms corresponding to three particle 
DA's of twist 3 and 4. Moreover, in exclusive 
amplitudes including three particle DA's of the 
$ \rho$-meson calculated in \cite{CZ} corresponding 
integrals contain logarithmic end-point divergences.
Our result (\ref{28}) then implies that if no special
cancellations occur (we see no reason why they 
should), in such a situation an attempt to evaluate
the twist 4 contribution will give rise to power
infrared divergences \footnote{Another possibility
(V. Braun, private communication) is that power 
infrared divergences may cancel by virtue of equations
of motion so that only logarithmic end-point 
singularities will survive. This scenario seems unlikely 
as the DA's (\ref{28}) are not related by the equations of 
motion. Dynamical calculations are needed to see whether  
cancellations of infrared divergences take place
in physical amplitudes.}
. This would mean a breakdown of OPE 
for this process and invalidation of a calculated  
twist 3 contribution. This observation can be of wider 
sense and imply that OPE may be inapplicable in
exclusive reactions beyond the leading twist 
order. This conjecture does not contradict to 
anything we currently understand about exclusive
processes. To illustrate this point, we consider two
examples. \\
 1. In the most well studied case of the 
pion electromagnetic form factor the short distance 
regime breaks down at the $ 1/Q^4 $ order as there
the soft Feynman-type mechanism \cite{BH}
brings in the same $ 1/Q^4 $ contribution as a twist
4 part of the hard rescattering amplitude (we leave
aside the mild Sudakov suppression in the
former). \\
2. For the 
$ \gamma \rho \pi $ transition form factor, which 
is a nonleading twist process \cite{CZ} related to 
the helicity flipping, the hard rescattering diagram
yields the asymptotic $ 1/Q^4 $ behavior. However, the 
soft Feynman contribution gives the same $ 1/Q^4 $
\cite{BH}, and therefore the short distance regime in this
process breaks down already at the leading order 
in $ 1/Q^2 $. If the above scenario were to occur, it 
would just signal on this breakdown, which is otherwise
not seen in the leading hard rescattering amplitude
taken on its own (logarithmic end-point
divergences are relatively harmless 
as they can be cured by the Sudakov 
correction factor). \\
 A common feature in these two examples
is that in the perturbative QCD approach the 
short distance regime is implied within the hard gluon
exchange mechanism. This {\it assumption} certainly
does not hold at the nonleading twist level 
as the $ 1/Q^4 $ parts of the hard amplitudes are just 
the radiative corrections to the soft Feynman-type 
contributions which have nothing to do with 
short distances. Thus it is not unreasonable to expect
divergences at the next order of OPE for these processes
as a manifestation of the same fact. It would
be challenging to see whether there
exist physical motivations to 
expect a breakdown of OPE in other and, in particular,
higher twist exclusive processes like
$ \psi \rightarrow \rho \pi $ which seems 
suggestive according to formulae given
in Ref.\cite{CZ}. To avoid possible
misunderstanding, we should stress that in the light 
cone QCD sum rule approach to exclusive processes 
\cite{BH} the 
situation is different as there the light cone 
dominance is enforced from outside by an analytic 
continuation to the deep Euclidean region. Therefore,
we do not expect a similar pattern of a breakdown of OPE
in this case, though this possibility cannot be 
ruled out.

\section*{Acknowledgments}  

I am grateful to A. Zhitnitsky for stimulating discussions.
Thanks are due to V. Braun for reading the manuscript 
and useful comments.

\end{document}